\documentclass[aps,prd,groupaddress,tighten,nofootinbib,superscriptaddress,twocolumn]{revtex4}
\usepackage{amssymb}
\usepackage{amsmath,color}
\usepackage{epsfig}
\usepackage{graphicx,epsf,epsfig}
\usepackage{bbm}
\def\slc#1{\setbox0=\hbox{$#1$}           % set a box for #1
    \dimen0=\wd0                                 % and get its size
    \setbox1=\hbox{/} \dimen1=\wd1               % get size of /
    \ifdim\dimen0>\dimen1                        % #1 is bigger
       \rlap{\hbox to \dimen0{\hfil/\hfil}}      % so center / in box
       #1                                        % and print #1
    \else                                        % / is bigger
       \rlap{\hbox to \dimen1{\hfil$#1$\hfil}}   % so center #1
       /                                         % and print /
    \fi}

\begin{document}
\vspace*{-2cm}
\begin{flushright}
NORDITA-2009-49, EURONU-WP6-09-07
\end{flushright}
%----------------------------------------------------------------------------------
\title{Non-standard neutrino interactions in the Zee--Babu model}
%----------------------------------------------------------------------------------
%\date{\today}
%----------------------------------------------------------------------------------
\author{Tommy Ohlsson}
\email{tommy_AT_theophys.kth.se}

\affiliation{Department of Theoretical Physics, School of
Engineering Sciences, Royal Institute of Technology (KTH) --
AlbaNova University Center, Roslagstullsbacken 21, 106 91 Stockholm,
Sweden}

\author{Thomas Schwetz}
\email{schwetz_AT_mpi-hd.mpg.de}

\affiliation{Max-Planck-Institut f{\"u}r Kernphysik, Postfach
103980, 69029 Heidelberg, Germany}

\author{He Zhang}
\email{zhanghe_AT_kth.se}

\affiliation{Department of Theoretical Physics, School of
Engineering Sciences, Royal Institute of Technology (KTH) --
AlbaNova University Center, Roslagstullsbacken 21, 106 91 Stockholm,
Sweden}
%----------------------------------------------------------------------------------
%\pacs{12.10.-g, 12.60.Jv, 14.60.Pq, 12.15.Ff}
%----------------------------------------------------------------------------------

\begin{abstract}
We investigate non-standard neutrino interactions (NSIs) in the
Zee--Babu model. The size of NSIs predicted by this model is
obtained from a full scan over the parameter space, taking into
account constraints from low-energy experiments such as searches for
lepton flavor violation (LFV) and the requirement to obtain a viable
neutrino mass matrix. The dependence on the scale of new physics as
well as on the type of the neutrino mass hierarchy is discussed. We
find that NSIs at the source of a future neutrino factory may be at
an observable level in the $\nu_e \to \nu_\tau$ and/or $\nu_\mu \to
\nu_\tau$ channels. In particular, if the doubly charged scalar of
the model has a mass in reach of the LHC and if the neutrino mass
hierarchy is inverted, a highly predictive scenario is obtained with
observable signals at the LHC, in upcoming neutrino oscillation
experiments, in LFV processes, and for NSIs at a neutrino factory.
\end{abstract}
\maketitle
%%%%%%%%%%%%%%%%%%%%%%%%%%%%%%%%%%%%%%%%%%%

\section{Introduction} \label{sec:intro}

Experimental studies of neutrino oscillations have provided us with
compelling evidence that neutrinos have masses and lepton flavors
mix. Among many possible mechanisms to describe the origin of
neutrino masses, radiative mass generation provides an attractive
method to obtain small neutrino masses. In such a framework,
neutrino masses are exactly vanishing at tree level, and are induced
as finite radiative corrections. Typically, neutrino masses are
suppressed by a loop factor and proportional to $\mu m_\ell^2/M^2$,
where $m_\ell$ are the charged lepton masses, $M$ is the mass scale of the
new particles in the loop, and $\mu$ is the scale of lepton-number
violation. Toghether with a modest suppression from Yukawa
couplings, this allows sub-eV neutrino masses, while having new
physics not too far from the electro-weak scale, $M \sim 1 \, {\rm
TeV}$, opening the possibility of collider tests of the neutrino
mass generation mechanism.

An economical way of radiative neutrino mass generation is to
enlarge the scalar sector of the Standard
Model~\cite{Konetschny:1977bn, Cheng:1980qt}. In the Zee model,
neutrino masses are obtained at one-loop level by adding a singly
charged scalar and a second $SU(2)_{\rm L}$ doublet to the Standard
Model~\cite{Zee:1980ai}. While the simplest version of this model
cannot accommodate current experimental data, since the predicted
leptonic mixing angle $\theta_{12}$ is too large (close to $\pi/4$),
a minor extension of the model remains a viable
option~\cite{Balaji:2001ex}. Alternatively, in the Zee--Babu
model~\cite{Zee:1985rj,Zee:1985id,Babu:1988ki}, two $SU(2)_{\rm L}$
singlet scalars are introduced, one singly and one doubly charged,
and neutrino masses are generated at two-loop level.
Phenomenological studies of this model have been performed, e.g., in
Refs.~\cite{Babu:2002uu, AristizabalSierra:2006gb, Nebot:2007bc}.
Through the exchange of heavy scalars, lepton flavor violating (LFV)
processes such as $\mu \to 3e$ and $\mu \to e\gamma$ can be
dramatically enhanced compared to the Standard Model. Furthermore,
the new scalars could be accessible for the Large Hadron Collider
(LHC). In particular, the doubly charged Higgs may induce very clean
like-sign bi-lepton events.

Besides colliders, next generation neutrino oscillation experiments
will also help us to unveil the underlying physics behind neutrino
masses. A neutrino factory will be the ultimate facility to perform
precision measurments of standard neutrino oscillations as well as
to search for non-standard neutrino properties. We take this as a
motivation to investigate non-standard neutrino interactions (NSIs)
in realistic neutrino mass models. In a given model, NSIs are
typically linked to LFV for charged leptons, yielding too tight
bounds, see, e.g., Refs.~\cite{Davidson:2003ha, Antusch:2008tz,
Biggio:2009nt}. For example, in the case of triplet scalar models
(i.e. type-II seesaw), NSIs are always entangled with the
interactions among four charged leptons, which suffer stringent
constraints from LFV processes like $\mu \to 3e$. There are no
sizable NSIs unless severe fine-tuning of Majorana phases is
invoked~\cite{Malinsky:2008qn}. In the case of the Zee--Babu model,
the situation is more involved, since the masses of singly and
doubly charged Higgs in principle can be well separated and a
different set of Yukawa couplings controls charged lepton and
neutrino interactions with the scalars. Non-trivial predictions for
the NSI parameters emerge from the different combinations of Yukawa
couplings responsible for the neutrino masses and mixing, and LFV
processes. In this respect, we will investigate NSIs in the
Zee--Babu model in detail.

The remaining parts of the work are organized as follows: In
Sec.~\ref{sec:model}, we sketch the Zee--Babu model and show how
NSIs are induced. In Sec.~\ref{sec:Constraints}, experimental
constraints from low-energy observables on the model parameters are
summarized, while in Sec.~\ref{sec:results}, we present the results
from our numerical study of NSI parameters within this model.
Discussion and summary follow in Sec.~\ref{sec:summary}.

%%%%%%%%%%%%%%%%%%%%%%%%%%%%%%%%%%%%%%%%%%%%%%%%%%%%%%%%%%%%%%%%%%%%%%%%%%%%%%%%%%%%%%
\section{The Zee--Babu model and non-standard neutrino interactions} \label{sec:model}
%%%%%%%%%%%%%%%%%%%%%%%%%%%%%%%%%%%%%%%%%%%%%%%%%%%%%%%%%%%%%%%%%%%%%%%%%%%%%%%%%%%%%%

In the minimal Zee--Babu model, two $SU(2)_{\rm L}$ singlet scalar
fields $h^+$ and $k^{++}$ are introduced with hypercharges $1$ and
$2$, respectively. The corresponding Lagrangian is then given by
\begin{eqnarray}\label{eq:L}
{\cal L} & = & {\cal L}_{\rm SM} + f_{\alpha\beta} L^{T}_{L\alpha} C
i\sigma_2 L_{L\beta} h^{+} + g_{\alpha\beta} \overline{e^c_\alpha}
e_{\beta} k^{++} \nonumber \\ &-& \mu h^- h^- k^{++} + {\rm h.c.} +
V_H \ ,
\end{eqnarray}
where $L_L$ denote left-handed lepton doublets, $e$ are the
right-handed charged leptons, and the scalar potential $V_H$
contains the couplings among scalar fields. The Yukawa couplings $f$
and $g$ are antisymmetric and symmetric, respectively. The trilinear
$\mu$ term breaks lepton number ($L$) in an explicit way, and hence,
one can naturally expect the dimensionful parameter $\mu$ to be
reasonably small, since the symmetry is enhanced in the limit $\mu
\to 0$.

Light neutrino masses are generated via a two-loop diagram, which
yields
\begin{eqnarray}\label{eq:m}
(m_{\nu})_{ab} = 16 \mu f_{ac} m_c g^*_{cd} I_{cd} m_d f_{bd} \ ,
\end{eqnarray}
where $m_c$ are charged lepton masses and $I_{cd}$ is a two-loop
integral \cite{McDonald:2003zj}. Since the $e^+e^-$ collider LEP at
CERN indicates that the masses of charged scalars are typically
larger than ${\cal O}(100~{\rm GeV})$, we can neglect the masses of
charged leptons compared to them. In this case, one finds
\begin{equation}\label{eq:I}
I_{cd} \approx I = \frac{1}{(16\pi)^2}\frac{1}{M^2}\frac{\pi^2}{3}
\tilde I \left(\frac{m_k^2}{m_h^2}\right) \,,
\end{equation}
where $M={\rm max}(m_k, m_h)$ and $\tilde I(r)$ is a dimensionless
function of order unity~\cite{Nebot:2007bc}. For our numerical
calculations we use the expression given in Eq.~(7) of
Ref.~\cite{Babu:2002uu}. Using Eq.~\eqref{eq:I}, the light neutrino
mass matrix becomes
\begin{eqnarray}\label{eq:mnu}
m_{\nu} \simeq \frac{1}{48\pi^2} \frac{\mu}{M^2} \, \tilde I \, f
D_e g^\dagger D_e f^T \ ,
\end{eqnarray}
where the matrix $D_e = {\rm diag}(m_e,m_\mu,m_\tau)$ contains the
charged-lepton masses. Light neutrino masses are suppressed by the
heavy scalar masses and proportional to the lepton-number violating
parameter $\mu$. Due to the antisymmetric property of $f$, we have
$\det m_\nu = 0$, and therefore, one of the neutrinos is massless if
higher-order corrections are not considered.

\begin{figure}[t]
\begin{center}\vspace{0.cm}
\includegraphics[width=8.5cm]{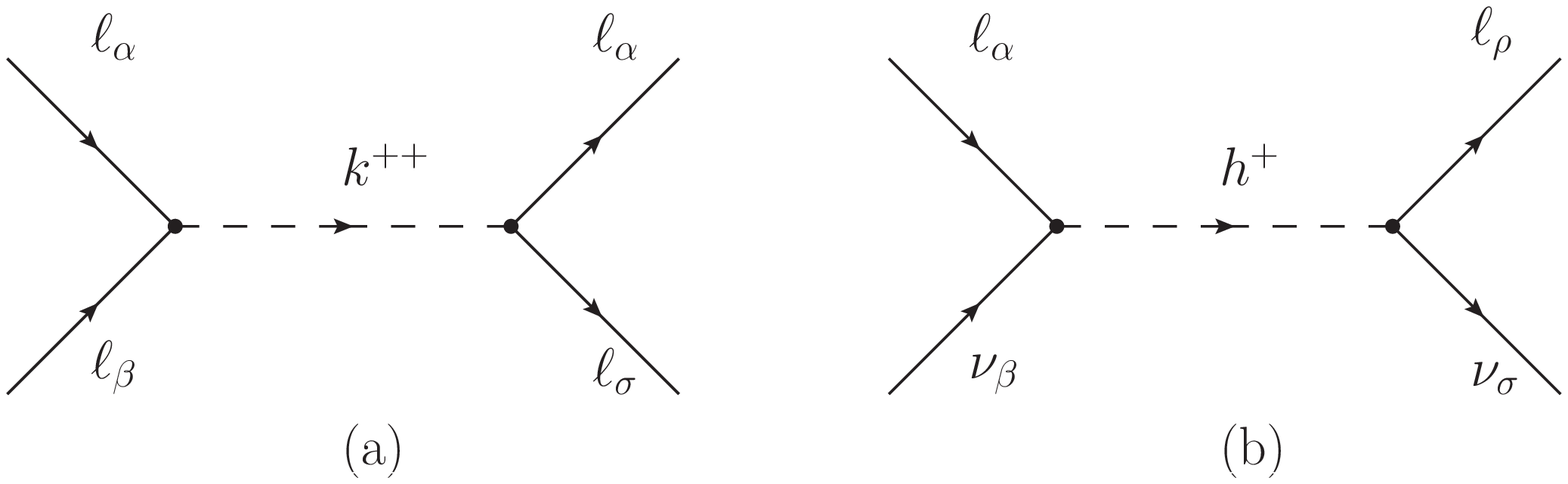}
\caption{\label{fig:fig1}Tree-level diagrams with the exchange of
heavy scalars. The corresponding diagrams are responsible for (a)
non-standard interactions of four charged lepton, and (b)
non-standard neutrino interactions.} \vspace{-0.cm}
\end{center}
\end{figure}

The heavy scalars will induce non-standard lepton interactions via
tree-level diagrams as shown in Fig.~\ref{fig:fig1}.
After integrating out the heavy scalars, the following dimension-6 operators
are generated at tree level~\cite{Antusch:2008tz}
\begin{eqnarray}
{\cal L}^{\rm NSI}_{d=6} & = & 4 \frac{f_{\alpha\beta}
f^*_{\rho\sigma}}{m^2_h} \left( \overline{\ell^c_\alpha} P_L
\nu_\beta \right) \left( \overline{\nu_\sigma} P_L \ell^c_\rho
\right) \nonumber \\ & = & 2 \frac{f_{\alpha\beta}
f^*_{\rho\sigma}}{m^2_h} \left( \overline{\ell_\rho} \gamma^\mu P_L
\ell_\alpha \right) \left( \overline{\nu_\sigma}\gamma_\mu P_L
\nu_\beta \right) \nonumber \\  & = & 2\sqrt{2}{G_F}
%\sum_{\footnotesize f,f' \atop C, \alpha,\beta}
\varepsilon^{\rho\sigma}_{\alpha\beta} \left(
\overline{\nu_\alpha}\gamma^\mu P_{L} \nu_\beta \right) \left(
\overline{\ell_\rho}\gamma_\mu P_{L} \ell_\sigma \right)\ ,
\end{eqnarray}
where a Fierz transformation has been applied for the second step,
and
\begin{eqnarray}\label{eq:defeps}
\varepsilon^{\rho\sigma}_{\alpha\beta} = \frac{f_{\sigma\beta}
f^*_{\rho\alpha}}{\sqrt{2} G_F m^2_h} \simeq 0.06 \, f_{\sigma\beta}
f^*_{\rho\alpha} \left(\frac {m_h}{\rm TeV}\right)^{-2}
\end{eqnarray}
are the canonical NSI parameters. For neutrinos propagating in normal
matter, only the following NSIs are induced
\begin{eqnarray}\label{eq:em}
\varepsilon^m_{\alpha\beta} = \varepsilon^{ee}_{\alpha\beta} = \frac{f_{e\beta}
f^*_{e\alpha}}{\sqrt{2} G_F m^2_h} \ .
\end{eqnarray}
Taking into account the antisymmetric property of $f$, one can find
that the only relevant NSI parameters in matter are
$\varepsilon^m_{\mu\tau}$, $\varepsilon^m_{\mu\mu}$, and
$\varepsilon^m_{\tau\tau}$. Furthermore, NSIs may show up at
neutrino production in a neutrino factory, related to the processes
$\mu\to e\overline{\nu_\beta}\nu_\alpha$ due to
$\varepsilon^{e\mu}_{\alpha\beta}$. To be consistent with the
notation in the literature, e.g., Ref.~\cite{Ohlsson:2008gx}, we
define
\begin{eqnarray}
  \varepsilon^s_{\mu\tau} &=&
  \varepsilon^{e\mu}_{\tau e} =
  \frac{f_{\mu e} f^*_{e\tau}}{\sqrt{2} G_F m^2_h} \ , \nonumber \\
  \varepsilon^s_{e\tau} &=&
  \varepsilon^{e\mu}_{\mu\tau} =
  \frac{f_{\mu\tau} f^*_{e\mu}}{\sqrt{2} G_F m^2_h} \ , \label{eq:es}
\end{eqnarray}
which correspond to the source effects in the $\nu_\mu \to \nu_\tau$
and $\nu_e \to \nu_\tau$ channels, respectively. By definition, the
relation
\begin{equation}\label{eq:mutau}
\varepsilon^m_{\mu\tau} = -\varepsilon^{s*}_{\mu\tau}
\end{equation}
holds, since both the NSI parameters $\varepsilon^m_{\mu\tau}$ and
$\varepsilon^s_{\mu\tau}$ are related to the Yukawa couplings $f_{e\mu}$ and
$f_{e\tau}$. Let us mention that Eq.~\eqref{eq:mutau} holds in a rather
general class of models, where NSIs are generated by dimension-6 operators
\cite{Gavela:2008ra}.

The light neutrino mass matrix can be diagonalized by means of a
unitary transformation as
\begin{eqnarray}
m_\nu = U D U^T \ ,
\end{eqnarray}
where $D={\rm diag}(m_1,m_2,m_3)$ and $U$ can be parametrized by
using three mixing angles and two CP-violating phases
\begin{widetext}
\begin{eqnarray}\label{eq:para}
U & = &\left(
\begin{matrix}c_{12} c_{13} & s_{12} c_{13} & s_{13}
e^{-{\rm i}\delta} \cr -s_{12} c_{23}-c_{12} s_{23} s_{13}
 e^{{\rm i} \delta} & c_{12} c_{23}-s_{12} s_{23} s_{13}
 e^{{\rm i} \delta} & s_{23} c_{13} \cr
 s_{12} s_{23}-c_{12} c_{23} s_{13}
 e^{{\rm i} \delta} & -c_{12} s_{23}-s_{12} c_{23} s_{13}
 e^{{\rm i} \delta} & c_{23} c_{13}\end{matrix}
\right) \left(
\begin{matrix} 1 & & \cr  & e^{{\rm i}\sigma} & \cr & & 1 \end{matrix}
\right) \ ,
\end{eqnarray}
\end{widetext}
where $c_{ij} \equiv \cos \theta_{ij}$, $s_{ij} \equiv \sin
\theta_{ij}$ (for $ij=12$, $13$, $23$), and $\delta$ is the Dirac
CP-violating phase. Here only one Majorana phase $\sigma$ is
involved, since one light neutrino is exactly massless. As a result
of $\det f =0$, there is an eigenvector
$v_0=(f_{\mu\tau},-f_{e\tau},f_{e\mu})$ which corresponds to the
zero eigenvalue $f v_0 =0$ \cite{Babu:2002uu}. Note that $v_0$ is
also an eigenvector of $m_\nu$, and therefore, we have
\begin{eqnarray}\label{eq:v0}
DU^Tv_0=0 \ ,
\end{eqnarray}
for both normal mass hierarchy ($m_1\ll m_2\ll m_3$, NH) and
inverted mass hierarchy ($m_2>m_1 \gg m_3$, IH).

Equation~\eqref{eq:v0} provides us with three equations, among which
one is trivially satisfied, since one element of $D$ is zero. The
other two equations lead to relations between $f$ and the lepton
mixing parameters. In the NH case, we have
\begin{eqnarray}\label{eq:eqNH1}
\frac{f_{e\tau}}{f_{\mu\tau}} & = &
\frac{s_{12}c_{23}}{c_{12}c_{13}} + \frac{s_{13}s_{23}}{c_{13}}
e^{-{\rm i}\delta} \ , \\
\label{eq:eqNH2}\frac{f_{e\mu}}{f_{\mu\tau}} & = &
\frac{s_{12}s_{23}}{c_{12}c_{13}} - \frac{s_{13}c_{23}}{c_{13}}
e^{-{\rm i}\delta} \ .
\end{eqnarray}
According to the current global fit of neutrino oscillation
experiments, the second terms in the right-hand sides of
Eqs.~\eqref{eq:eqNH1} and \eqref{eq:eqNH2} can be neglected, since
they are suppressed by the small mixing angle $\theta_{13}$. Then,
we obtain the approximate relation $f_{e\mu} \simeq f_{e\tau}\simeq
f_{\mu\tau}/2$. Taking into account the experimental constraints
from Eqs.~\eqref{eq:Univ} and \eqref{eq:LFV} below, we can roughly
estimate that $|f_{e\mu}| \sim |f_{e\tau}| \lesssim 0.05~(m_h/{\rm
TeV})$ and $|f_{\mu\tau}| \lesssim 0.1~(m_h/{\rm TeV})$. Compared
with Eqs.~\eqref{eq:em} and \eqref{eq:es}, the possibly important
NSI parameter is $\varepsilon^s_{e\tau}$, which mainly affects the
$\nu_e \to \nu_\tau$ channel. The CP-violating phases of
$f_{\alpha\beta}$ are suppressed by $\theta_{13}$, and thus, NSIs
cannot induce very distinctive CP-violating effects in the NH case.

For the IH case, the two non-trivial equations are
\begin{eqnarray}\label{eq:eqIH1}
\frac{f_{e\tau}}{f_{\mu\tau}} & = &-\frac{s_{23}c_{13}}{s_{13}}
e^{-{\rm i}\delta} \ , \\
\label{eq:eqIH2} \frac{f_{e\mu}}{f_{\mu\tau}} & = &
\frac{c_{13}c_{23}}{s_{13}} e^{-{\rm i}\delta} \ .
\end{eqnarray}
Indeed, it is obvious that $|f_{e\mu}| \sim |f_{e\tau}|$ and
$|f_{\mu\tau}| \sim |f_{e\tau}| s_{13}/s_{23} $ hold. Thus, the
potentially sizable NSI parameters are $\varepsilon^m_{\mu\tau}$,
$\varepsilon^m_{\mu\mu}$, $\varepsilon^m_{\tau\tau}$ for neutrino
propagation in matter, and $\varepsilon^s_{\mu \tau}$ for source
effects in the $\nu_\mu \to \nu_\tau$ channel at a neutrino factory.
Equations~\eqref{eq:eqIH1} and \eqref{eq:eqIH2} imply that
$\varepsilon^s_{\mu\tau}$ and $\varepsilon^m_{\mu\tau}$ are real,
whereas the phase of $\varepsilon^s_{e\tau}$ is given by $\delta$.
This may lead to an interesting correlation of CP-violation in
standard oscillations and $\varepsilon^s_{e\tau}$-induced
CP-violating effects~\cite{GonzalezGarcia:2001mp,
FernandezMartinez:2007ms}.

%%%%%%%%%%%%%%%%%%%%%%%%%%%%%%%%%%%%%%%%%%%%%%%%%%%%%%%%%%%%%%%%%%
\section{Experimental constraints} \label{sec:Constraints}
%%%%%%%%%%%%%%%%%%%%%%%%%%%%%%%%%%%%%%%%%%%%%%%%%%%%%%%%%%%%%%%%%%

At low-energy scales, stringent constraints from LFV processes
mediated by the heavy scalars have to be included when we confront
the model with experimental data. In the following, we compile the
bounds given in Ref.~\cite{Nebot:2007bc}.
\begin{itemize}
\item $\ell^-_a \to \ell^+_b \ell^-_c \ell^-_d$. As shown in Fig.~\ref{fig:fig1},
these rare lepton decays are mediated by $k^{++}$ at tree level, and
set very stringent constraints on the corresponding Yukawa coupling
$g$ and the mass of $k^{++}$. The bounds at 90~\% C.L.\ read
\begin{eqnarray}\label{eq:Rare}
|g_{e\mu} g^*_{ee}| & < & 2.3 \times 10^{-5}~(m_k/{\rm TeV})^2 \ , \nonumber \\
|g_{e\tau} g^*_{ee}| & < & 0.010~(m_k/{\rm TeV})^2 \ , \nonumber \\
|g_{e\tau} g^*_{e\mu}| & < & 0.006~(m_k/{\rm TeV})^2 \ , \nonumber \\
|g_{e\tau} g^*_{\mu\mu}| & < & 0.008~(m_k/{\rm TeV})^2 \ , \nonumber \\
|g_{\mu\tau} g^*_{ee}| & < & 0.008~(m_k/{\rm TeV})^2 \ , \nonumber \\
|g_{\mu\tau} g^*_{e\mu}| & < & 0.008~(m_k/{\rm TeV})^2 \ , \nonumber \\
|g_{\mu\tau} g^*_{\mu\mu}| & < & 0.010~(m_k/{\rm TeV})^2 \ .
\end{eqnarray}
Note that NSIs are induced by exchanging the singly charged Higgs
$h^+$, and one can in principle tune the mass of $k^{++}$ or the
scale of $g$ in order to suppress its contributions to the LFV
decays. However, since all the parameters $f, g, m_h$, and $m_k$
enter in the expression for the neutrino mass matrix, c.f.\
Eq.~(\ref{eq:mnu}), the constraints from Eqs.~(\ref{eq:Rare}) have
to be included also in an analysis of NSIs, in order to obtain the
correct parameter space available for the model.

\item $\mu^+ e^- \to \mu^- e^+$. The muonium to antimuonium
conversion through the exchange of $k^{++}$ are well bounded
experimentally. For the similar reason mentioned above, these
constraints are mainly on $m_k$ and $g$, and the current bound at
90~\% C.L.\ is $|g_{ee} g^*_{\mu\mu}| < 0.2~(m_k/{\rm TeV})^2$.

\item Universality in $\ell_a \to \ell_b \overline{\nu} \nu$ decays.
The Fermi coupling constant measured in muon and tau decays
obtains corrections from the exchange of $h^+$, which sets strong
constraints on the Yukawa coupling $f$:
\begin{eqnarray}\label{eq:Univ}
|f_{e\mu}|^2 & < & 0.015~(m_h/{\rm TeV})^2 \ , \nonumber \\
\left||f_{\mu\tau}|^2 - |f_{e\tau}|^2 \right| & < & 0.05~(m_h/{\rm TeV})^2 \ , \nonumber \\
\left||f_{e\tau}|^2 - |f_{e\mu}|^2 \right| & < & 0.06~(m_h/{\rm TeV})^2 \ , \nonumber \\
\left||f_{\mu\tau}|^2 - |f_{e\mu}|^2 \right| & < & 0.06~(m_h/{\rm
TeV})^2 \ .
\end{eqnarray}

\item Rare lepton decays $\ell^-_\alpha \to \ell^-_\beta \gamma$.
Both $h^{+}$ and $k^{++}$ contribute to LFV photon interactions at
one-loop level, and the most stringent bound comes from $\mu\to
e\gamma$. Neglecting the contributions from the doubly charged
Higgs, we obtain experimental bounds at 90~\% C.L.
\begin{eqnarray}\label{eq:LFV}
|f^*_{e\tau}f_{\mu\tau}|^2 & < & 3.4\times 10^{-5}~(m_h/{\rm TeV})^4 \ , \nonumber \\
|f^*_{e\mu}f_{\mu\tau}|^2 & < & 1.7 ~(m_h/{\rm TeV})^4 \ , \nonumber \\
|f^*_{e\mu}f_{e\tau}|^2 & < & 0.7 ~(m_h/{\rm TeV})^4 \ .
\end{eqnarray}
Note that in our numerical analysis, contributions from both singly
and doubly charged Higgs are included, see, e.g.,
Ref.~\cite{Nebot:2007bc}.
\end{itemize}

Besides the bounds above, there also exist other constraints, like
the $\mu-e$ conversion in nuclei and the anomalous magnetic moments
of the muon, which are relatively loose, and hence will not be
considered in this work. The perturbativity of the model imposes
limits on the Yukawa couplings $f$ and $g$, in particular for very
massive charged scalars. Similarly, the stability of the vacuum
requires $\mu \ll 4\pi \min(m_k,m_h)$~\cite{Babu:2002uu}.

%%%%%%%%%%%%%%%%%%%%%%%%%%%%%%%%%%%%%%%%%%%%%%%%%%%%%%%%%%%%%%%%%%%%%%%%%%%%%
\section{Numerical results for NSI in the Zee--Babu model}\label{sec:results}
%%%%%%%%%%%%%%%%%%%%%%%%%%%%%%%%%%%%%%%%%%%%%%%%%%%%%%%%%%%%%%%%%%%%%%%%%%%%%

We have performed a full scan of the parameter space of the model in
order to obtain predictions for NSIs. Following
Refs.~\cite{Babu:2002uu, Nebot:2007bc}, we take as independent
parameters the lepton mixing angles, the Dirac and Majorana phases
$\delta$ and $\sigma$, the Yukawa couplings $g_{ee}$, $g_{e\mu}$,
$g_{e\tau}$ and one of the three $f_{\alpha\beta}$'s, and the scalar
masses $m_h$, $m_k$, as well as the trilinear coupling $\mu$.
(Neutrino mass-squared differences are fixed to their best-fit values,
since their uncertainties are comparably small.) The remaining Yukawa
couplings $f_{\alpha\beta}$ and $g_{\alpha\beta}$ are then fixed by
Eqs.~\eqref{eq:mnu}, \eqref{eq:eqNH1}, and \eqref{eq:eqNH2} for NH or
Eqs.~\eqref{eq:mnu}, \eqref{eq:eqIH1}, and \eqref{eq:eqIH2} for
IH. For each set of these parameters, we compare the model predictions
to the experimental data with a $\chi^2$ function
\begin{eqnarray}\label{eq:chi2}
\chi^2 = \sum_i \frac{(\rho_i - \rho^0_i)^2}{\sigma^2_{i}} \ ,
\end{eqnarray}
where $\rho^0_i$ represents the data of the $i$th experimental observable,
$\sigma_i$ the corresponding 1$\sigma$ absolute error, and $\rho_i$ the
prediction of the model. The experimental observables are the neutrino
mixing angles (taken from Ref.~\cite{Schwetz:2008er}), and the
constraints from LFV and universality tests given in
Eqs.~(\ref{eq:Rare})--(\ref{eq:LFV}).

For the dimensionful parameters $m_h, m_k$, and $\mu$, we adopt
first two representative choices, namely $m_h=m_k=\mu=10 \, {\rm
TeV}$ or $m_h=m_k=\mu=1\, {\rm TeV}$. The latter case might be just
in reach for the LHC. For a luminosity of 300~fb$^{-1}$ and under
optimistic assumptions, this may lead to order 10 four-lepton events
from the pair-production of the doubly charged scalars $pp \to
k^{++}k^{--} \to \ell^+\ell^+\ell^-\ell^-$~\cite{Nebot:2007bc}. We
do not consider much lower scalar masses, since already at 1~TeV
many of the experimental bounds are saturated. At 10~TeV, the
constraints are less tight, which leaves more freedom in choosing
the parameters of the model, with the obvious disadvantage of not
being testable at LHC.
Choosing $\mu$ of the same order as the scalar masses is
conservative in the following sense. For fixed scalar masses,
neutrino masses are proportional to $\mu$. Hence, decreasing $\mu$
would require to increase the Yukawa couplings, see
Eq.~\eqref{eq:mnu}. This would make the constraints from
Sec.~\ref{sec:Constraints} more severe and the parameter space would
be more constrained. Therefore, we decided to take $m_h=m_k=\mu$ in
order to keep $\mu$ relatively large, but still ensure the stability
of the vacuum~\cite{Babu:2002uu}. Towards the end of this section we
also investigate the case $m_h \neq m_k$.

\begin{figure*}
\begin{center}\vspace{-0.7cm}
  \includegraphics[width=8cm,bb=0 0 750 750]{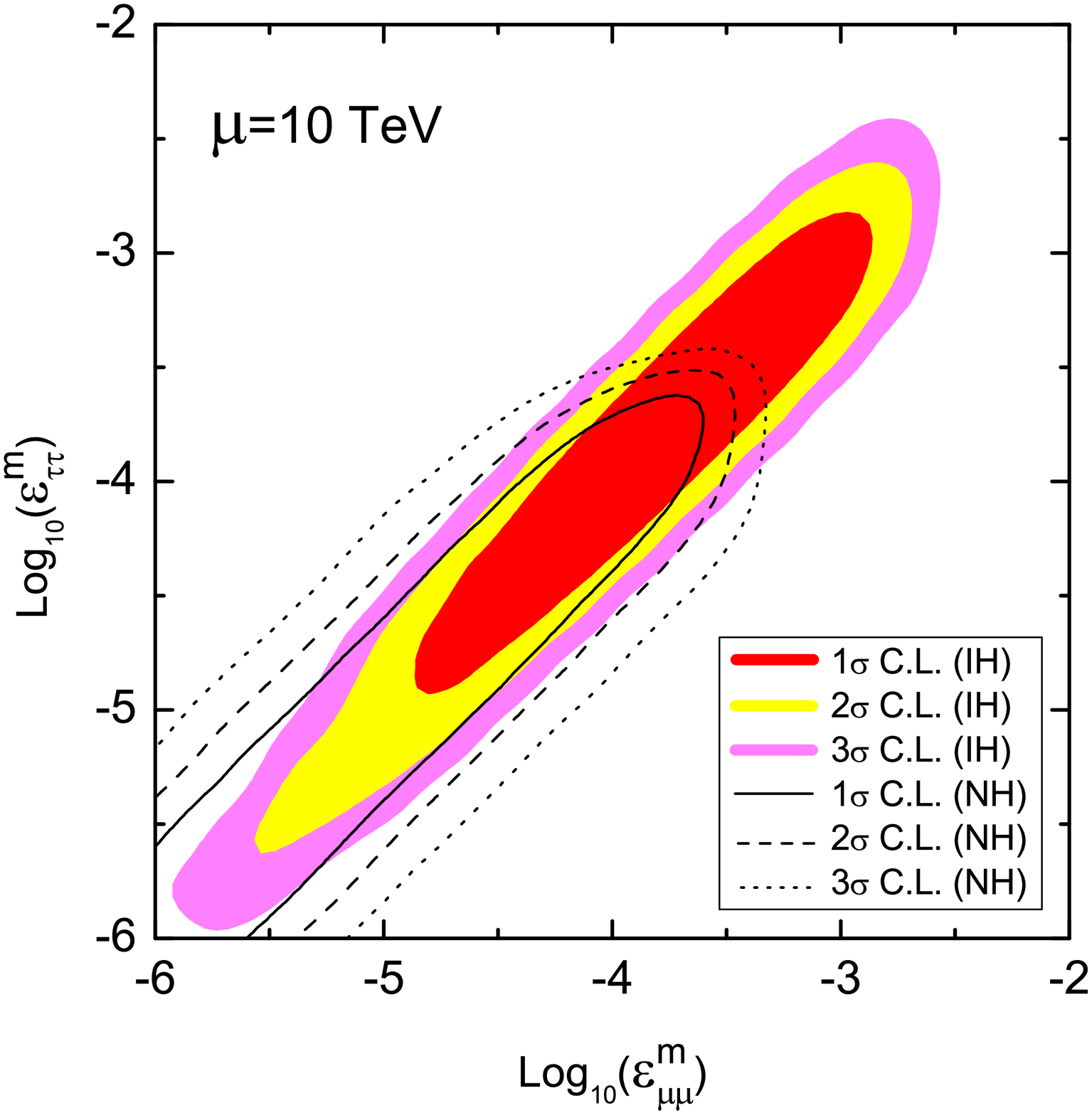}
  \includegraphics[width=8cm,bb=0 0 750 750]{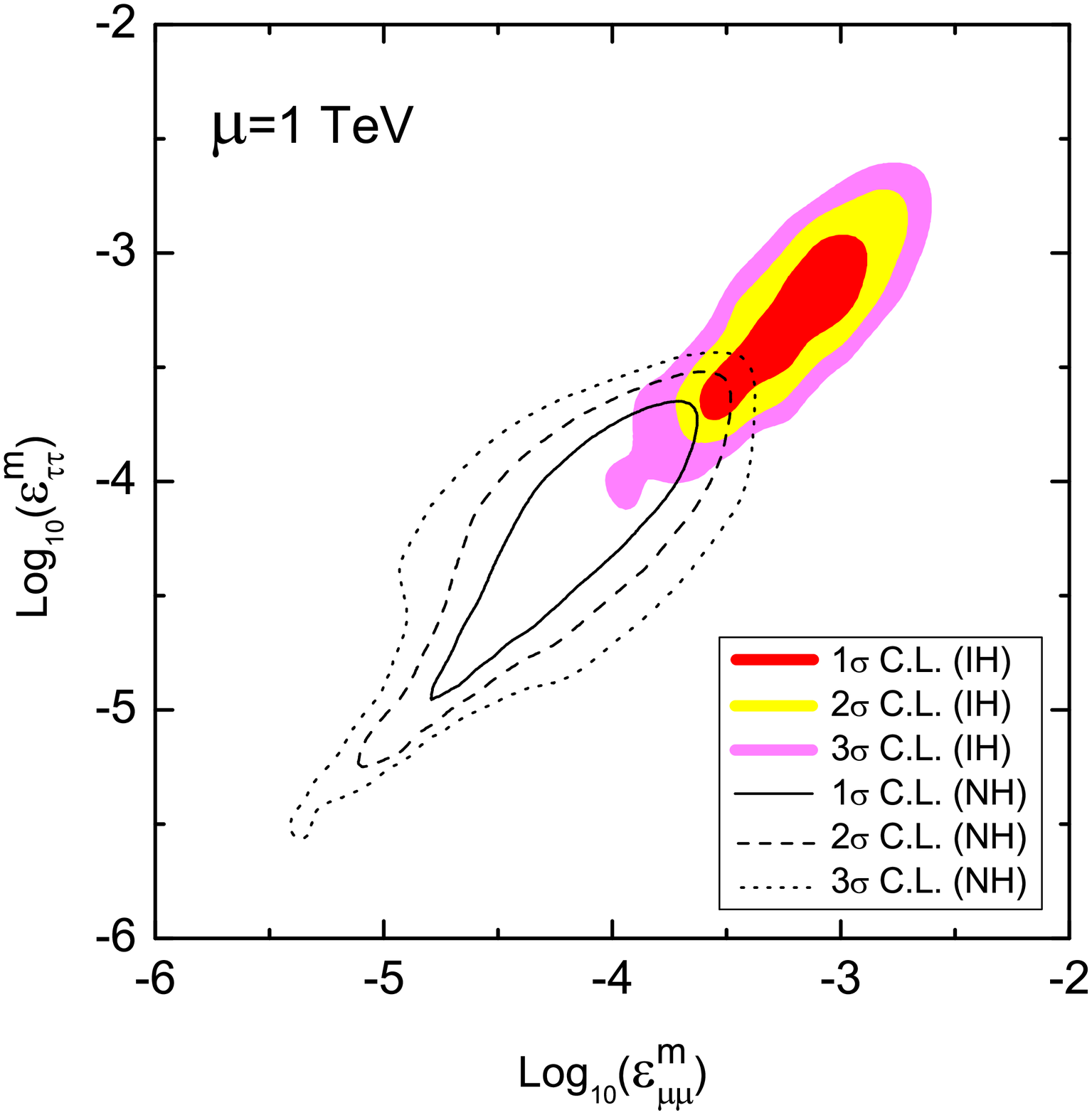}
  \includegraphics[width=8cm,bb=0 -90 750 660]{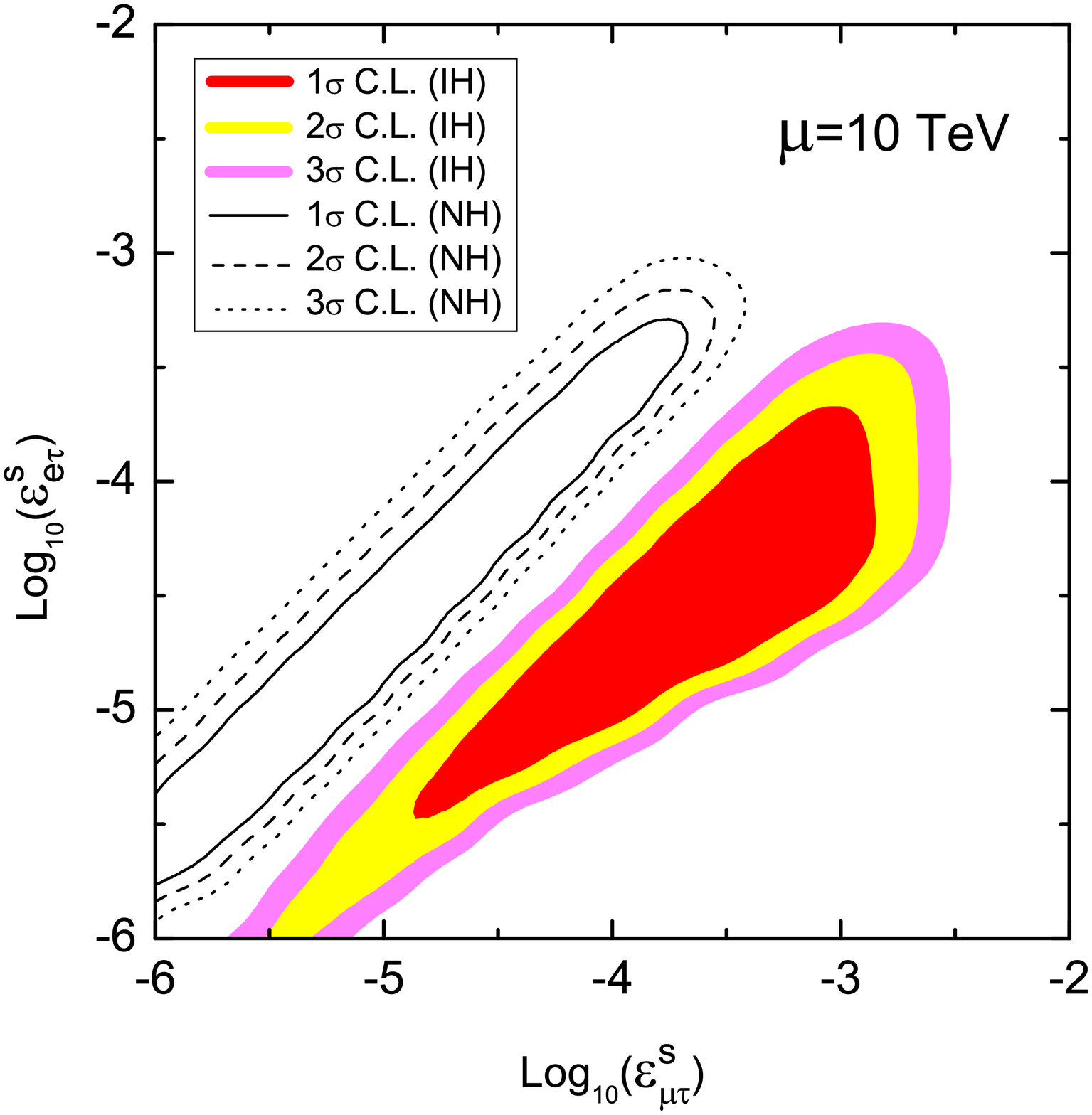}
  \includegraphics[width=8cm,bb=0 -90 750 660]{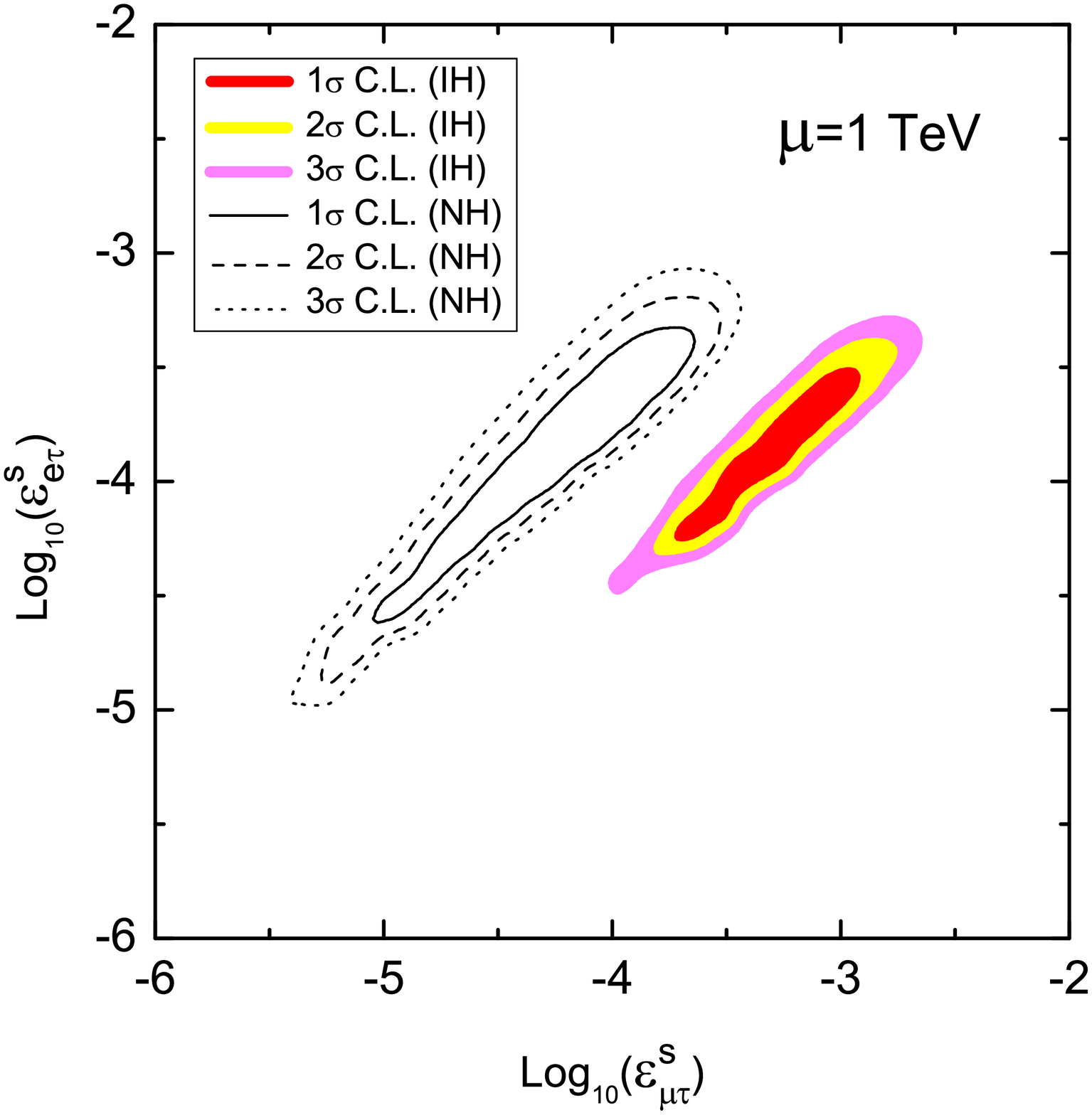}
\vspace{-0.7cm}
  \caption{\label{fig:fig2} The allowed region of NSI parameters at
  1$\sigma$, 2$\sigma$, and 3$\sigma$ C.L.\ in the Zee--Babu model. We take
  $m_h=m_k=\mu=10~{\rm TeV}$ for the figures in the left-hand side column and
  $m_h=m_k=\mu=1~{\rm TeV}$ for the figures in the right-hand side column.}
\end{center}
\end{figure*}

In Fig.~\ref{fig:fig2}, we present the allowed regions of NSI
parameters at 1$\sigma$, 2$\sigma$, and 3$\sigma$ C.L., defined as
contours in $\Delta\chi^2$ for two degrees of freedom with respect
to the $\chi^2$ minimum. The left (right) panels correspond to
scalar masses of 10~TeV (1~TeV).
The upper plots in Fig.~\ref{fig:fig2} show the NSI parameters
$|\varepsilon^m_{\mu\mu}|$ and $|\varepsilon^m_{\tau\tau}|$ for
neutrino propagation. We find that values of order $\varepsilon \sim
10^{-3}$ can be obtained only in the IH case, whereas in the NH case
they are typically one order of magnitude smaller. These flavor
diagonal propagation NSI parameters induce effectively a
non-standard matter effect and it has been shown that present
long-baseline experiments are not very sensitive to these NSI
effects~\cite{Meloni:2009ia}. Even a two-baseline neutrino factory
would only be sensitive to such NSI parameters at the level of
$\varepsilon\gtrsim 10^{-2}$~\cite{Kopp:2008ds}. The lower plots in
Fig.~\ref{fig:fig2} indicate that the NSI parameter
$|\varepsilon^s_{e\tau}|$ relevant at the source of a neutrino
factory may reach values up to $10^{-3}$ for both hierarchies. In
the IH case, the parameters $|\varepsilon^m_{\mu\tau}| =
|\varepsilon^s_{\mu\tau}|$ may also be as large as few $\times
10^{-3}$. For a scalar mass scale of 1~TeV, a non-trivial lower
bound on the NSI parameters of order $10^{-4}$ is found in the
right-hand column of Fig.~\ref{fig:fig2}. Indeed, for scalar masses
in the TeV range, the model is rather constrained and the
requirement of a correct neutrino mass matrix pushes the Yukawa
couplings close to the bounds from Sec.~\ref{sec:Constraints}
\cite{AristizabalSierra:2006gb, Nebot:2007bc}, which in turn implies
``large'' NSIs. The extended region seen in the figure comes mainly
from the freedom to adjust the Dirac CP-violating phase $\delta$.

Source NSIs related to the muon decay in a neutrino factory are probed
efficiently with a near detector, see e.g., Ref.~\cite{Datta:2000ci}, since
they lead to the appearance of ``wrong'' neutrino flavors even at ``zero
distance'' \cite{Antusch:2006vwa}. For the cases of interest in the
Zee--Babu model, $|\varepsilon^s_{e\tau}|$ and $|\varepsilon^s_{\mu\tau}|$,
obviously a tau detector at the near site would be useful
\cite{FernandezMartinez:2007ms, Malinsky:2008qn, Malinsky:2009gw,
Tang:2009na, Antusch:2009pm, Malinsky:2009df}. The authors of
Ref.~\cite{Tang:2009na} consider as an example a 2~kt OPERA-like near
detector and find sensitivities for $|\varepsilon^s_{e\tau}|,
|\varepsilon^s_{\mu\tau}| \gtrsim 7\times 10^{-4}$. Note that in order to
disentangle the effect of $\varepsilon^s_{e\tau}$ and
$\varepsilon^s_{\mu\tau}$, the ability to identify the charge of the tau
lepton would be required.

%%%%%%%%%%%%%%%%%%%%%%%%%%%%%%%%%%%%%
\begin{figure*}[t]
\begin{center}\vspace{-0.7cm}
\includegraphics[width=8cm,bb=0 0 750 750]{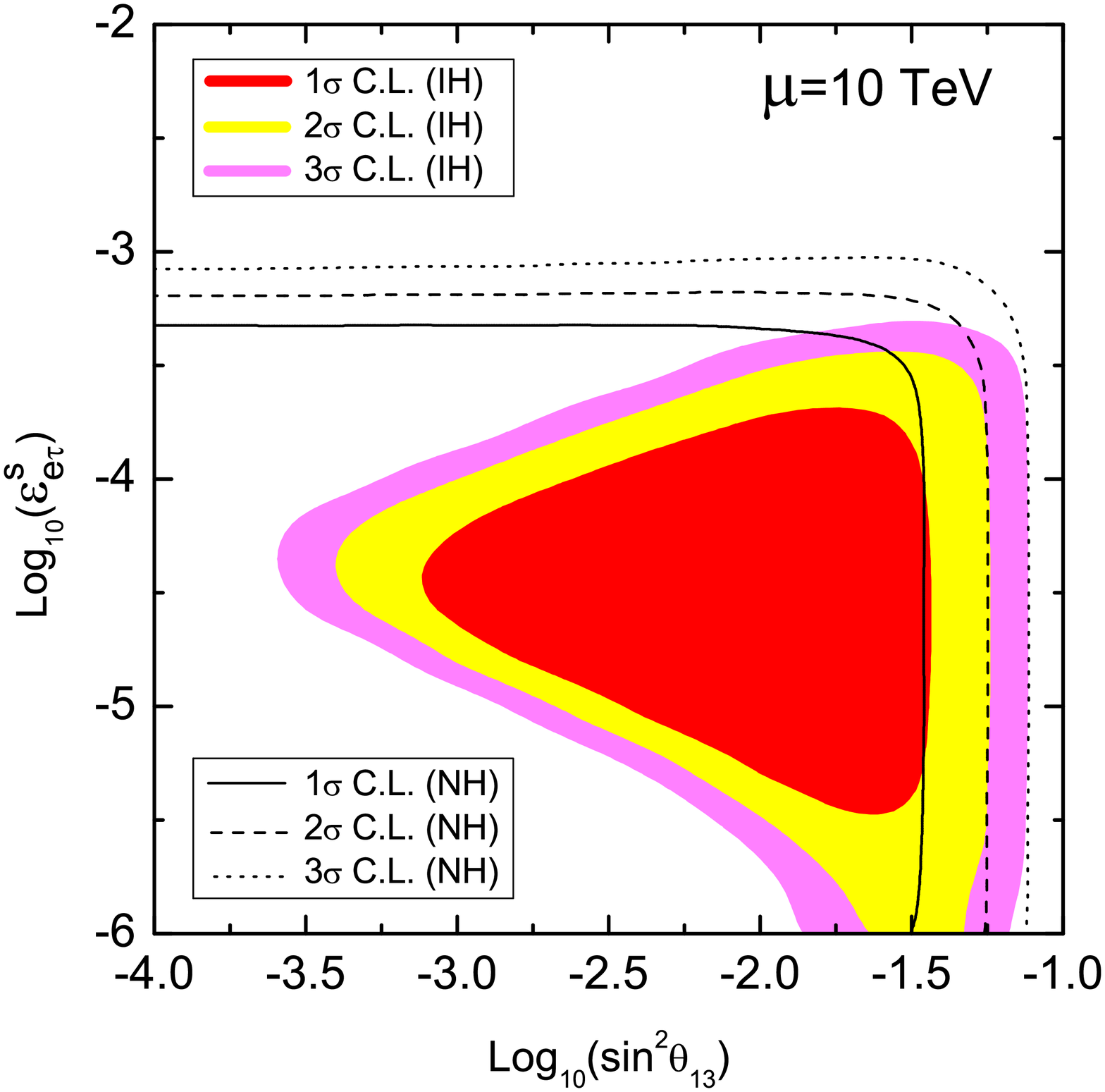}
\includegraphics[width=8cm,bb=0 0 750 750]{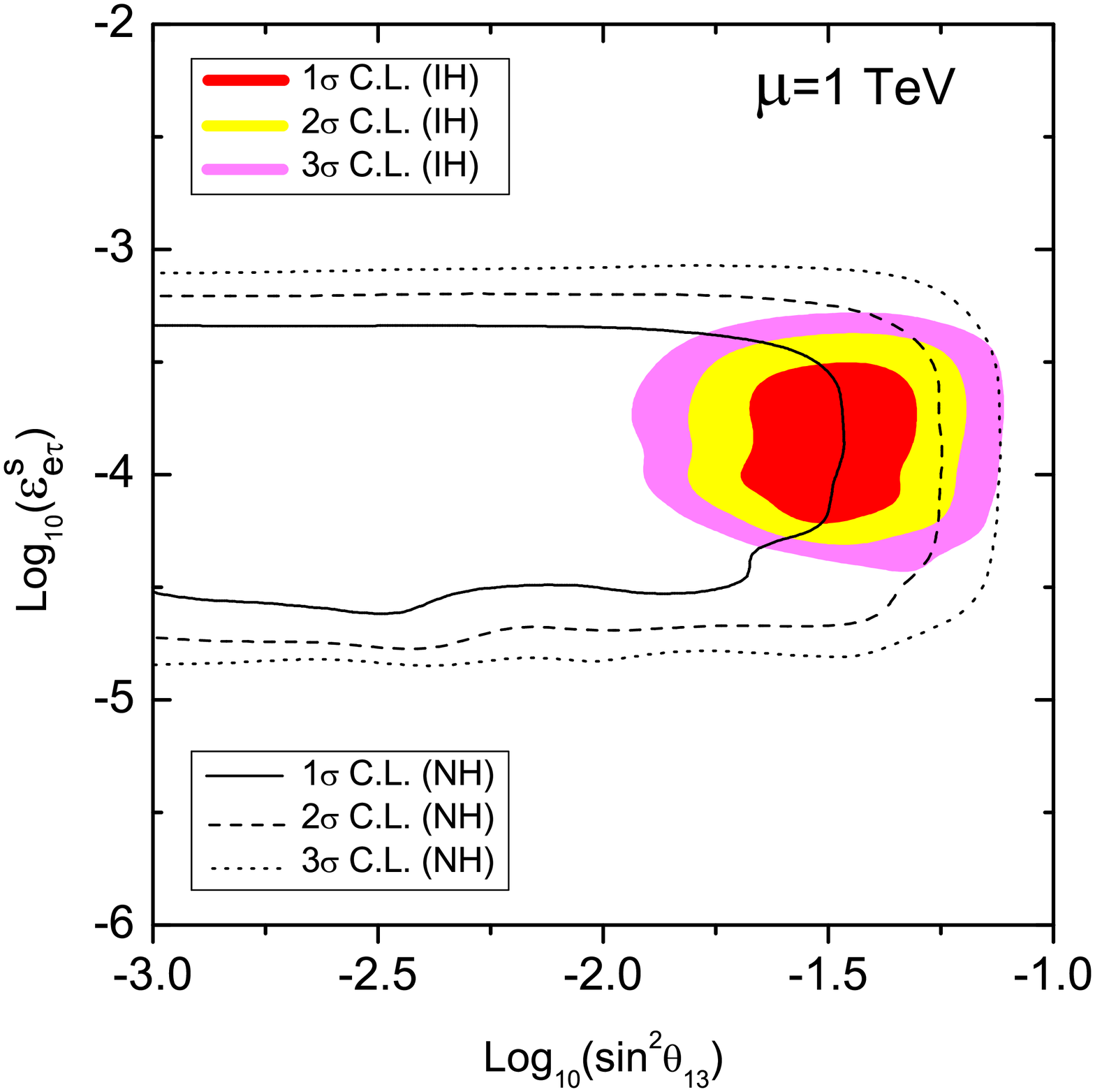}
\includegraphics[width=8cm,bb=0 -90 750 660]{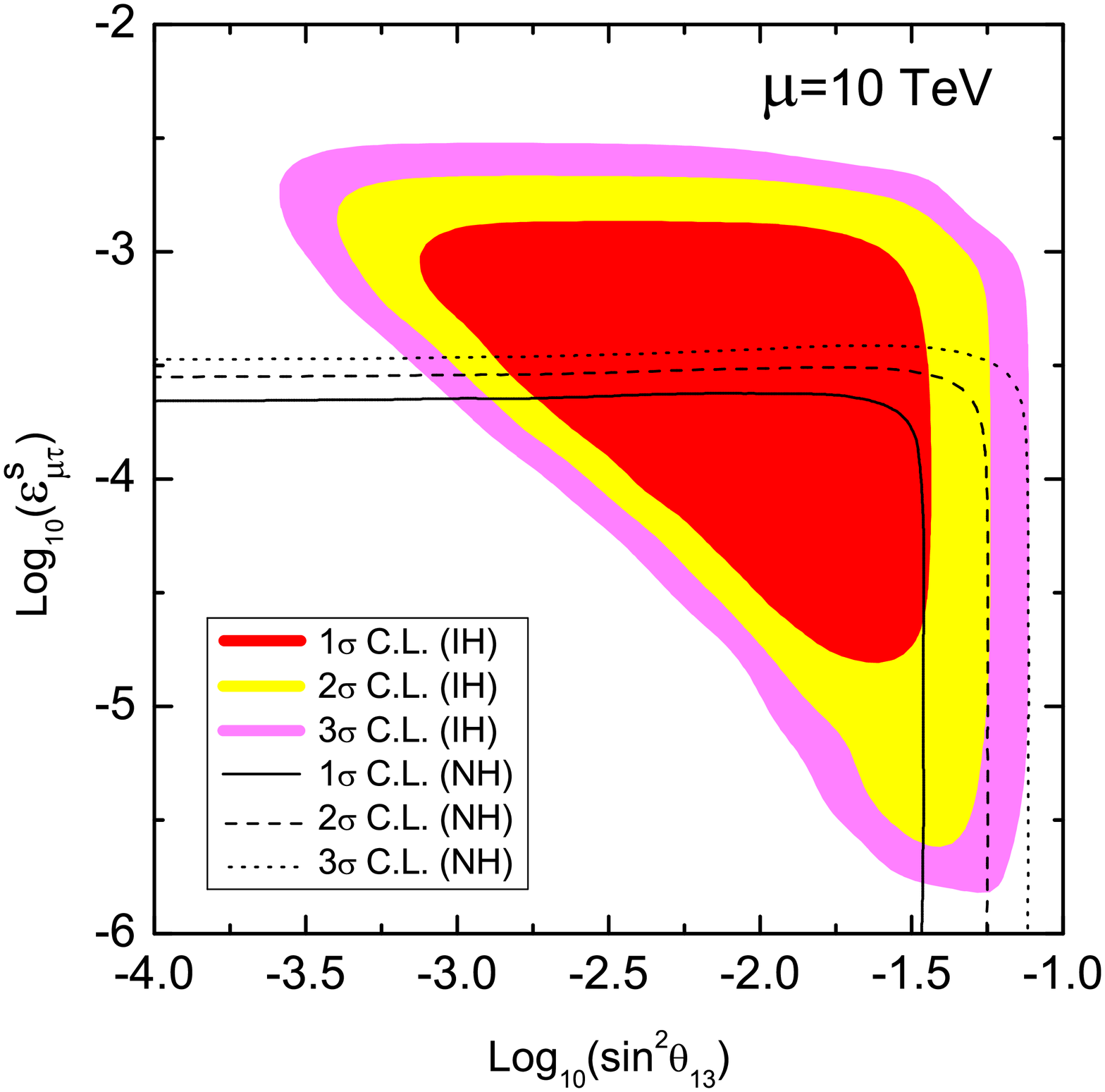}
\includegraphics[width=8cm,bb=0 -90 750 660]{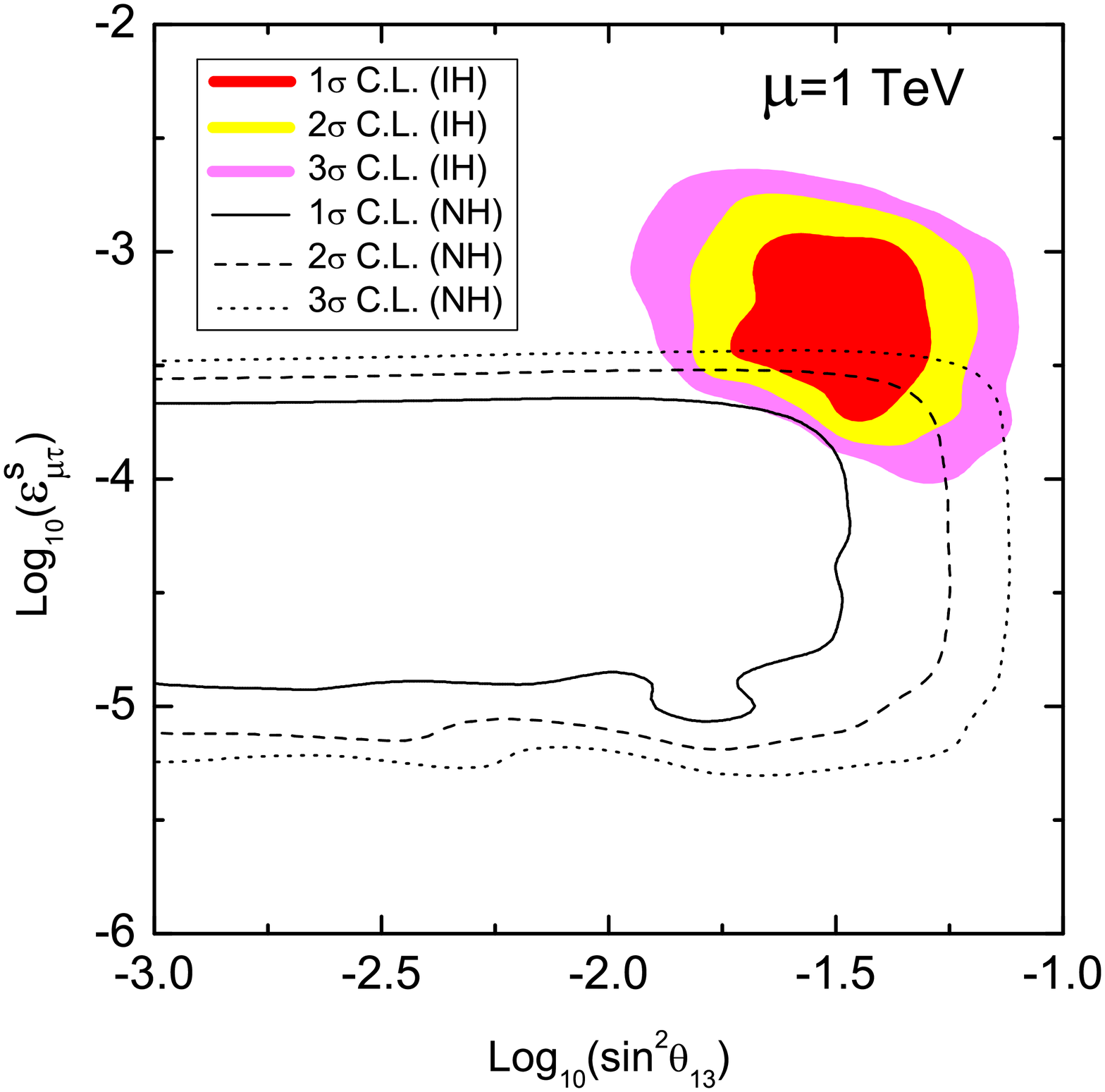}
  \vspace{-0.7cm} \caption{\label{fig:theta} Allowed regions at 1$\sigma$,
  2$\sigma$, and 3$\sigma$ C.L.\ in the plane of $\sin^2\theta_{13}$ and
  $|\varepsilon^s_{e\tau}|$ (upper panels) or $|\varepsilon^s_{\mu\tau}| =
  |\varepsilon^m_{\mu\tau}|$ (lower panels).  We take $m_h=m_k=\mu=10~{\rm TeV}$
  for the figures in the left-hand side column and $m_h=m_k=\mu=1~{\rm TeV}$ for
  the figures in the right-hand side column.}
\end{center}
\end{figure*}
%%%%%%%%%%%%%%%%%%%%%%%%%%%%%%%%%%%%%

The sensitivity to $|\varepsilon^m_{\mu\tau}|$ for neutrino
propagation has been discussed for atmospheric
neutrinos~\cite{Fornengo:2001pm, GonzalezGarcia:2004wg}, the OPERA
long-baseline experiment~\cite{Blennow:2008ym}, superbeam
experiments~\cite{Kopp:2007ne}, and a neutrino factory, e.g.\ in
Refs.~\cite{Huber:2001zw, Goswami:2008mi, Kopp:2008ds,
Antusch:2009pm}. Typically, the reach is at a few $\times 10^{-2}$
or worse, which is not sufficient to probe the parameter range
predicted by the Zee--Babu model. However, there are two reasons why
in our case we may expect better prospects to observe NSI effects in
this channel. First, in many of the above mentioned studies the
complex phase of $\varepsilon^m_{\mu\tau}$ has been marginalized
over, whereas in the Zee--Babu model $\varepsilon^m_{\mu\tau}$ is
predicted to be real, see the discussion after Eqs.~\eqref{eq:eqIH1}
and \eqref{eq:eqIH2}. Second, Eq.~\eqref{eq:mutau} relates source
and propagation NSIs in this channel. The relevance of the phases
can be understood from Eq.~(35) of Ref.~\cite{Kopp:2007ne}, which
shows that the relevant leading terms in the survival probability
$P_{\mu\mu}$ are proportional to
$|\varepsilon^s_{\mu\tau}|\sin(\phi^s)$ and
$|\varepsilon^m_{\mu\tau}|\cos(\phi^m)$, where $\phi^{s,m} \equiv
{\rm arg}(\varepsilon^{s,m}_{\mu\tau})$. Hence, these terms can be
set to zero if $\phi^s$ and $\phi^m$ can be chosen independently,
but if they are coupled by $\phi^s = \pi-\phi^m$ following from
Eq.~\eqref{eq:mutau}, at least one of them will always be non-zero.
Under these special conditions, we estimate from the results of
Refs.~\cite{Antusch:2009pm, Tang:2009na} sensitivities of a neutrino
factory for $\varepsilon^m_{\mu\tau}$ in the range of $10^{-3}$,
even without a near detector. In the presence of a near
tau-detector, a sensitivity for $|\varepsilon^m_{\mu\tau}| =
|\varepsilon^s_{\mu\tau}| > 6\times 10^{-4}$ is
reached~\cite{Tang:2009na}.

In Fig.~\ref{fig:theta}, we show the correlations between NSI
parameters and the mixing angle $\theta_{13}$. In the IH case and
scalar masses at the TeV scale, one obtains a quite strong
prediction for the mixing angle $\theta_{13}$. From
Eqs.~\eqref{eq:eqIH1} and \eqref{eq:eqIH2} follows that
$|f_{\mu\tau}|$ is suppressed by $s_{13}$, whereas a correct
neutrino mass matrix requires $f_{\mu\tau}$ to be of the same order
as $f_{e\mu}$ and $f_{e\tau}$. Figure~\ref{fig:theta} indeed shows
that for the IH case and scalar masses in the TeV range, values of
$\theta_{13}$ close to its present bound are
predicted~\cite{Nebot:2007bc}, with a lower bound of
$\sin^2\theta_{13} \gtrsim 10^{-2}$. Such a sizable lower bound is
of particular interest, since it would guarantee a discovery at the
forthcoming reactor \cite{Guo:2007ug,Ardellier:2006mn} or
long-baseline \cite{Itow:2001ee} experiments in the near future
\cite{Huber:2009cw}.

In the NH case, no lower bound on $\theta_{13}$ is obtained. In this
case, the presence of NSIs may have an impact on the search for
$\theta_{13}$ at a future neutrino factory, especially if
$\theta_{13}$ is relatively small. In particular,
$\varepsilon^s_{e\tau}$ may lead to $\nu_\tau$ at the source, which
will oscillate to $\nu_\mu$ and lead to so-called ``wrong-sign''
muons in the far detector, which might be confused with the effect
of a tiny $\theta_{13}$~\cite{Huber:2002bi}.

\begin{figure*}[t]
\begin{center}\vspace{-0.7cm}
\includegraphics[width=8cm,bb=0 0 750 750]{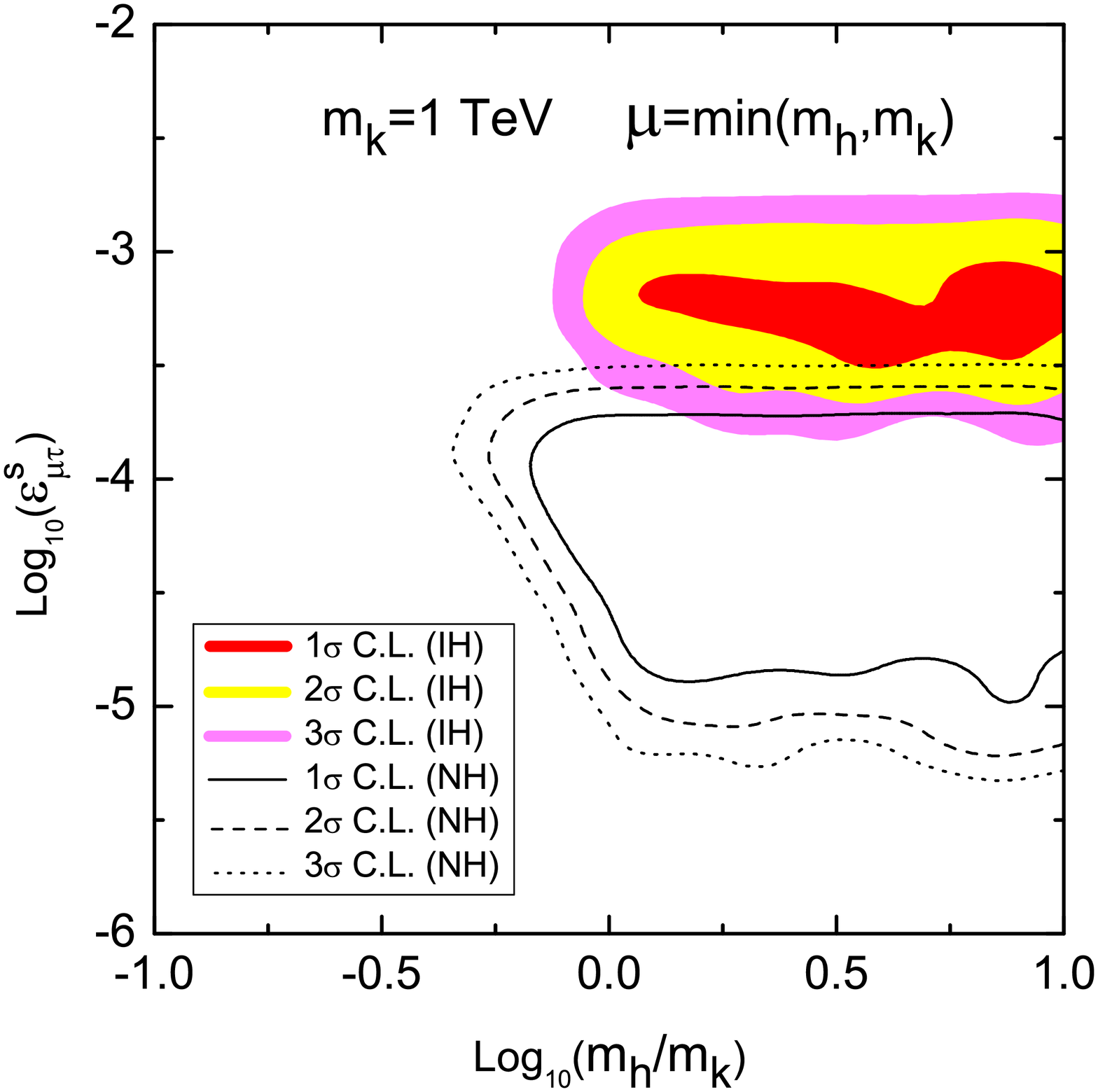}
\includegraphics[width=8cm,bb=0 0 750 750]{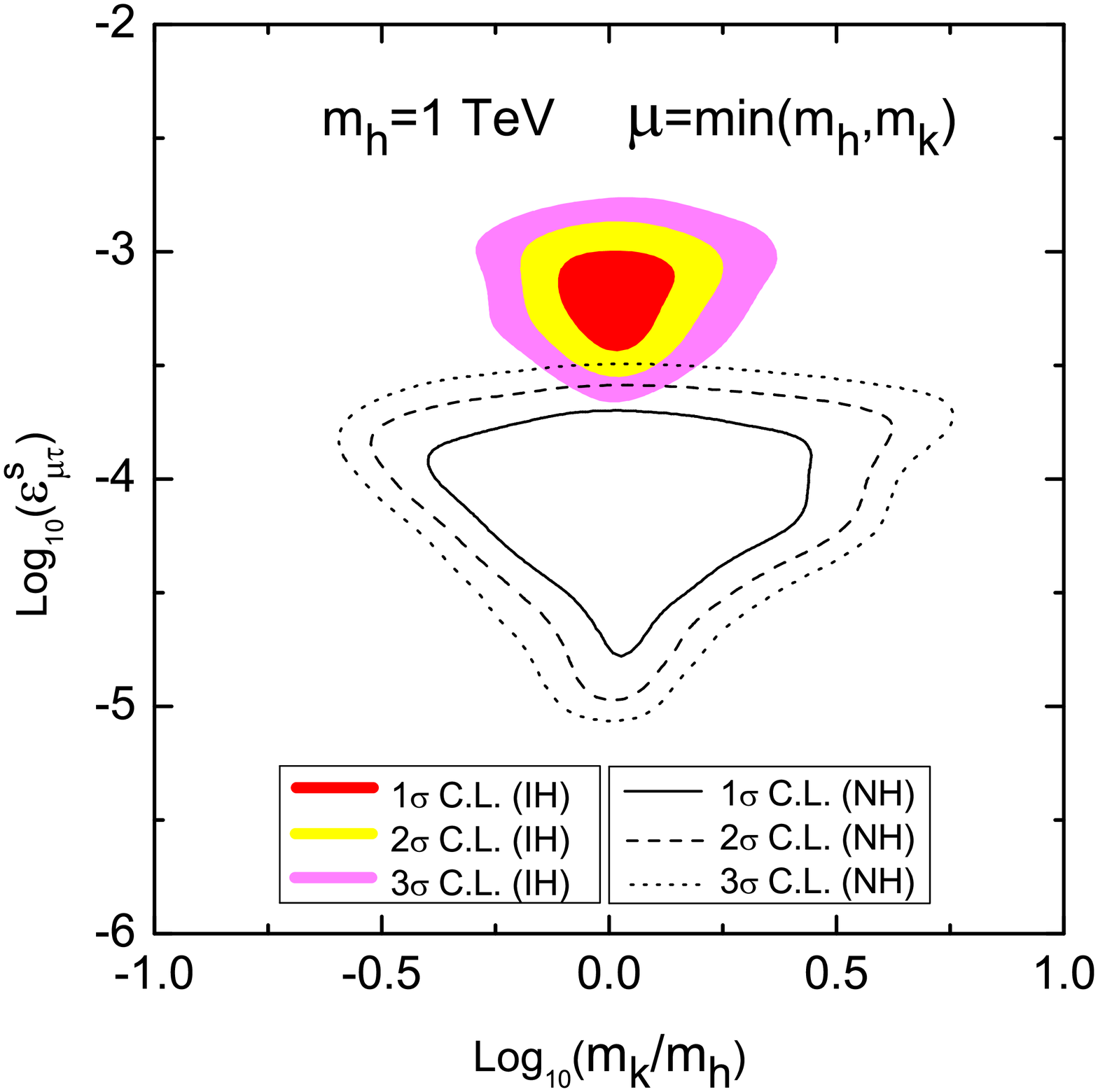}
  \caption{\label{fig:scalars} Allowed regions at 1$\sigma$, 2$\sigma$, and
  3$\sigma$ C.L.\ in the plane of the scalar masses and
  $|\varepsilon^s_{\mu\tau}| = |\varepsilon^m_{\mu\tau}|$. In the left
  panel,
  we vary the mass of the singly charged scalar $m_h$ and fix the doubly
  charged one at $m_k = 1$~TeV, while in the right panel, we vary $m_k$ and
  fix $m_h = 1$~TeV. In both cases, we use $\mu = {\rm min}(m_h, m_k)$.}
\end{center}
\end{figure*}

Finally, let us relax the assumption $m_h = m_k$ and investigate the
dependence on the masses of the scalars. In Fig.~\ref{fig:scalars},
we show the size of NSI parameters by fixing one of the two scalars
at 1~TeV and varying the mass of the other one.
%%%%%%%%%%%%%%%%%%%%%%%%%%%%%%%%%%%%%
%%%%%%%%%%%%%%%%%%%%%%%%%%%%%%%%%%%%%
In order to guarantee the stability of the
vacuum~\cite{Babu:2002uu}, while keeping neutrino masses as large as
possible, we use for the scale of lepton number violation $\mu =
{\rm min}(m_h, m_k)$. The qualitative behaviour of these results can
be understood from the expression for the neutrino mass matrix in
Eq.~\eqref{eq:mnu} and the fact that the constraints on the Yukawa
couplings $f$ and $g$ from Sec.~\ref{sec:Constraints} scale with
$m_{h}$ and $m_k$, respectively.

First, the lower bound on the scalar masses follows from the fact
that decreasing either $m_h$ or $m_k$ decreases the neutrino mass
because of $\mu = {\rm min}(m_h, m_k)$. At the same time, the bounds
on the Yukawas become more severe and it is impossible to obtain
sufficiently large neutrino masses. Second, if we increase $m_k$,
while keeping $m_h$ at 1~TeV (right panel of
Fig.~\ref{fig:scalars}), neutrino masses decrease with $m_k^{-2}$
because of $M = {\rm max}(m_h, m_k)$. Since the constraints on $g$
increase only with $m_k$, it is not possible to compensate for the
$m_k^{-2}$ suppression by increasing $g$. Hence, if the singly
charged scalar is at the TeV scale also the doubly charged one has
to be in that range. However, the opposite statement is not true and
relatively large $m_h$ is possible for $m_k = 1$~TeV. If $m_h$ is
increased, again neutrino masses decrease as $m_h^{-2}$, but in this
case there are two factors of $f$ entering Eq.~\eqref{eq:mnu}, and
since the constraints on them increase as $m_h$, it is possible to
keep neutrino masses constant by increasing $f$. However, note that
this does not lead to larger NSIs, since $\varepsilon \propto
f^2/m_h^2$, see Eq.~\eqref{eq:defeps}, which is constant, in
agreement with the left panel of Fig.~\ref{fig:scalars}.

We conclude that from the point of view of NSIs, the crucial mass
parameter is the one of the doubly charged scalar. Note also that
this one has the most striking signature at colliders, namely the
decay into two like-sign leptons. Apparently, vastly separated
masses for the singly and doubly charged scalars is either not
phenomenologically viable or does not affect the prediction for NSIs.
Therefore, the NSI results obtained for $m_h=m_k$ are generic.

%%%%%%%%%%%%%%%%%%%%%%%%%%%%%%%%%%%%%%%%%%%%%%%%%%%%%%
\section{Discussion and summary} \label{sec:summary}
%%%%%%%%%%%%%%%%%%%%%%%%%%%%%%%%%%%%%%%%%%%%%%%%%%%%%%

We have studied NSIs in the Zee--Babu two-loop neutrino mass model,
which are mediated by the singly charged scalar of the model. We
have shown that non-standard neutrino matter interactions relevant
for the propagation of neutrinos may be induced. The relevant
parameters $\varepsilon^m_{\mu\mu}$, $\varepsilon^m_{\tau\tau}$, and
$\varepsilon^m_{\mu\tau}$ may reach values of order $10^{-3}$. While
flavor diagonal NSIs of this size are too small to be observable,
the off-diagonal term $\varepsilon^m_{\mu\tau}$ may be within the
reach of a future neutrino factory. In addition to these matter
NSIs, NSIs affecting the muon decay at the source of a neutrino
factory may be induced in the Zee--Babu model, in both the $\nu_e
\to \nu_\tau$ ($\varepsilon^s_{e\tau}$) and $\nu_\mu \to \nu_\tau$
($\varepsilon^s_{\mu\tau}$) channels.\footnote{Note that in
superbeam or reactor experiments the neutrino production proceeds
via hadronic interactions, which are not affected in this model.}
The possible size of NSI parameters depends on the scale of new
physics (i.e., the masses of the singly and doubly charged scalars,
$m_h$ and $m_k$, respectively, and the scale of the lepton-number
violating parameter $\mu$) and on the type of the neutrino mass
hierarchy, NH or IH.

The most constrained situation is obtained for IH and scalar masses
at 1~TeV. In this case, the NSI parameters $\varepsilon^s_{e\tau}$
and $\varepsilon^s_{\mu\tau}$ are predicted to be in the range
$10^{-4}-10^{-3}$, probably in reach of a near tau-detector at a
future neutrino factory~\cite{Tang:2009na}. Thanks to the fact that
$\varepsilon^s_{\mu\tau}$ is real in this model combined with the
relation $\varepsilon^s_{\mu\tau} = -\varepsilon^{m*}_{\mu\tau}$,
even a standard neutrino factory without a near tau-detector may be
sensitive to the NSI values predicted in this case. Furthermore,
this configuration predicts a value of $\theta_{13}$ in reach of the
upcoming oscillation experiments \cite{Huber:2009cw}, as well as
signals in LFV processes close to the present bounds, with good
prospects for a signal in $\mu \to
e\gamma$~\cite{AristizabalSierra:2006gb, Nebot:2007bc}.

If kinematically accessible, the singly and doubly charged scalars of
the model could be directly produced through the $s$-channel
processes at the Tevatron and the LHC. There is no severe
suppression of the cross section for the production of doubly
charged scalar $k^{++}$, and if $2m_{h^+} > m_{k^{++}}$, it will
predominantly decay into like-sign charged-lepton pairs with a very
striking experimental signature. However, note that doubly charged
scalars occur in a variety of models (e.g., the triplet scalar model
for neutrino mass), and therefore, complementary signatures are
required to identify the model. If the singly charged and doubly
charged scalars are found at LHC, then---besides signals in LFV
searches---the Zee--Babu model predicts a rather large value of
$\theta_{13}$ and signals for NSI at a neutrino factory at a level
of $10^{-3}$ if the mass hierarchy is inverted. In case of NH,
$\theta_{13}$ as well as NSIs may still be in reach of future
experiments, but no signal is guaranteed in either case. A similar
situation emerges if the scale of new physics is beyond the reach of
the LHC, for example at 10~TeV. In this case, observable signals may
still arise for NSIs at a neutrino factory (as well as for
$\theta_{13}$), but no relevant lower bound is obtained.

In conclusion, we have shown that, in the case of the Zee--Babu
model for radiative neutrino masses, the interplay of the
phenomenology at colliders, searches for LFV, and NSI effects at a
neutrino factory could play a complementary role towards the goal of
identifying the true mechanism of neutrino mass generation.

\begin{acknowledgments}

\vspace{-2.5mm} We wish to thank Walter Winter and Joachim Kopp for
useful discussions. We acknowledge the hospitality and support from
the NORDITA scientific program ``Astroparticle Physics
--- A Pathfinder to New Physics'', March 30 -- April 30, 2009 during
which parts of this study was performed. This work was supported by
the Royal Swedish Academy of Sciences (KVA) [T.O.], the G{\"o}ran
Gustafsson Foundation [H.Z.], and the Swedish Research Council
(Vetenskapsr{\aa}det), contract no.~621-2008-4210 [T.O.]. T.S.\
acknowledges support of the Transregio SFB TR27 ``Neutrinos and
Beyond'' der Deutschen Forschungsgemeinschaft and by the EC under
the European Commission FP7 Design Study: EUROnu, Project Nr.\
212372. The EC is not liable for any use that may be made of the
information contained herein.

\end{acknowledgments}

%\bibliography{bib-babu}

\end{document}